*Women in science: overview, challenges, and inspirations*

Débora Peres Menezes
Departamento de Física, CFM, Universidade Federal de Santa Catarina, Florianópolis, SC, Brasil
Diretoria de Análise de Resultados e Soluções Digitais, CNPq, Brasília, DF, Brasil
 debora.p.m.26@gmail.com, debora.menezes@cnpq.br

**Abstract**
     The purpose of this article is to contextualise the current situation of gender inequality in the Brazilian society and, in particular, in science, and to show actions that try to minimise this situation and support girls who want to become scientists and women who are already in their careers. Historical and current data on the Brazilian reality are highlighted and show that social mishaps cannot be ignored and that the barriers faced by female science students, especially in the so-called STEM areas (hard sciences, technology, engineering and maths), can be a source of dropout. Some specific awards for female researchers are mentioned and projects that aim to show the challenges of the career and inspire future scientists are listed. The lack of gender equality must be a source of recurring discussions in academic environments and the obstacles faced by under-represented groups must be remembered and confronted if we want to move towards a fairer and more diverse Brazilian science.

**Mulheres na ciência: panorama, desafios e inspirações**

**Resumo**
     Este artigo tem o propósito de contextualizar a atual situação de desigualdade de gênero na sociedade brasileira e, em especial, na ciência e mostrar ações que tentam minimizar esse quadro e apoiar meninas que desejem se tornar cientistas e mulheres que já estão na carreira. Dados históricos e atuais sobre a realidade brasileira são salientados e mostram que os percalços sociais não podem ser ignorados e que as barreiras enfrentadas por estudantes de ciência do gênero feminino, especialmente nas áreas ditas STEM (ciências exatas, tecnologia, engenharia e matemática), podem ser fonte de desistência. Alguns prêmios específicos para pesquisadoras são lembrados e projetos que têm a finalidade de mostrar os desafios da carreira e inspirar futuras cientistas são elencados. A falta de equidade de gênero deve ser assunto recorrente nos ambientes acadêmicos e o enfrentamento dos obstáculos por grupos sub-representados precisa ser lembrado e confrontado, se quisermos caminhar para uma ciência brasileira mais justa e diversa.

**Palavras-chave:** estereótipo de gênero, efeito tesoura, vieses inconscientes, mídias sociais, mulheres na ciência

**1. Introdução**
     O Conto de Aia, livro de Margareth Atwood [1], é uma obra de ficção distópica, mas o que se passou no Afeganistão, condenando mulheres antes livres para escolher o que vestir e estudar a se recolher dentro de burcas, é real. O Brasil tem trilhado um caminho

que o aproxima de práticas religiosas fundamentalistas muito perigosas. Em junho de 2024, um projeto apresentado na Câmara dos Deputados [2], que ficou conhecido como o PL do estuprador, previa penas para aborto legal, incluindo casos de estupro, maiores do que as penas existentes para os estupradores. Após clamor popular, ele foi retirado de pauta, mas retornou em novembro numa forma ainda mais cruel, como PEC [3], sigla para proposta de emenda constitucional e foi aprovado na Comissão da Constituição e Justiça (CCJ) da Câmara Federal, num sinal de que os corpos das mulheres brasileiras correm o risco de ser usados pela extrema direita como barganha política.

Não bastasse esse quadro bastante assustador, o movimento masculinista também cresce no Brasil. Com fóruns de discussão que vêm ganhando força na DeepWeb e aliciando meninos, o movimento prega o retorno da soberania masculina por meio de práticas muito violentas, como o estupro e tortura de mulheres. Um dos braços mais conhecidos do masculinismo é o movimento *Red Pill,* nome que faz alusão ao filme Matrix, no qual a pílula vermelha dava acesso ao conhecimento real dos fatos. Movimentos anteriores, como o *Backlash,* que procurava desacreditar argumentos científicos sobre igualdade de gênero, sucumbiram, mas num momento de grande disseminação de discursos de ódio pelas redes sociais, o crescimento do masculinismo preocupa estudiosos de violência de gênero nas plataformas digitais [4].

Este preâmbulo é para nos lembrar que parece não haver direitos garantidos para as mulheres, que precisam se mobilizar constantemente para assegurá-los. Mas como combater o fundamentalismo religioso e o conservadorismo de pessoas que entendem que os corpos e as vontades das mulheres podem ser propriedade alheia? Um dos modos é educá-las para garantir que cheguem a espaços de liderança e de poder. E a ciência é um desses espaços.

Neste artigo, vou mencionar dados sobre desigualdade de gênero, discorrer sobre ações em andamento que visam disseminar informação qualificada sobre o assunto e dar visibilidade ao papel das mulheres e, por fim, apresentar o projeto Mulheres na Ciência, que venho coordenando desde 2019.

## 2. A realidade dos dados

Infelizmente, nas áreas conhecidas como STEM (*Science, Technology, Engineering and Mathematics*), da qual faz parte a física, a presença das mulheres é muito pequena e é fruto de uma construção social que permeia as suas vidas há muito tempo. A primeira Lei Educacional do Império, promulgada em 15 de outubro de 1827 [5] alijava as meninas do ensino da matemática, uma vez que, para um dos parlamentares mais influentes da época, o Visconde de Cayru, as meninas só precisavam aprender o básico, uma vez que "Deus deu barbas aos homens" [6]. Essa alegação é tão bizarra que me faz pensar que ele entendia que as pessoas pensavam com as barbas e não com os neurônios…

Essa construção social existe até hoje e fica patente nas barreiras e micro-agressões diárias que as mulheres precisam enfrentar, como as mencionadas no início deste texto. Estudos mostram que o estereótipo de gênero afeta a autoimagem das meninas desde muito cedo. Enquanto aos 5 anos, as meninas identificam pessoas muito inteligentes ao próprio sexo, aos 7, as meninas já se enxergam como menos inteligentes que os meninos [7]. Num estudo que ouviu estudantes, pais e professores em São Paulo, Buenos Aires e Cidade do México [8], 90% das meninas entrevistadas alegou que 'engenharia é coisa de menino'. Além do estereótipo de gênero, há outras barreiras mais sutis, os vieses inconscientes. Um estudo desse efeito mostra a escolha enviesada por um currículo masculino, mesmo quando idêntico a outro feminino [9]. Ainda há os famosos casos de mulheres inviabilizadas, como as três protagonistas do filme Estrelas Além do Tempo [10], Katherine Johnson, Dorothy Vaughan e Mary Jackson e o emblemático efeito tesoura, que corta as mulheres de posição de destaque e de poder [11]. Exemplos clássicos são as laureadas com prêmio Nobel: 65 mulheres e 908 homens até 2024, sendo apenas 5 na física (dentre 229), 3 na economia (dentre 96) e 8 na química (dentre 200). Casos mais próximos da nossa realidade são a existência de apenas 1 mulher Ministra de Ciência e tecnologia (Luciana Santos), 1 mulher presidente da Academia Brasileira de Ciências, instituição que existe desde 1916 (Helena Nader) e 3 mulheres no Supremo Tribunal Federal, instituído em 1889, com a promulgação da República.

Volto agora minha atenção para dados educacionais. Segundo dados do Censo da educação Superior feito pelo Inep em 2019 (antes da pandemia) [12], o Brasil formou 1,2 milhão de pessoas, sendo que 124,4 mil se formaram em pedagogia, 121,2 mil em direito e 91,9 mil em administração de empresas, isto é, essas 3 áreas, todas de humanidades formaram 30% de todos os graduados no país. Nesse mesmo ano, o Brasil formou 365 matemáticos e, para cada matemático formado, o Brasil formou 10 teólogos. Também em 2019, o Brasil formou 2.400 professores de física e quase 30.000 professores de educação física, isto é, para cada prof. de física, há mais de 10 professores de educação física, num retrato que faz parecer que as academias de ginástica têm mais valor no Brasil de hoje do que o ensino de ciência nas escolas. Além disso, estima-se que a cada 4 estudantes que ingressam no curso de física, 3 desistem, evadem. Em 1 ano, os EUA formam 8x mais astrônomos do que o Brasil forma em uma década! E certamente, esses números pioraram com a pandemia.

Na Folha de São Paulo de 20/09/2022, Marcelo Viana, o atual Diretor geral do Instituto de Matemática Pura e Aplicada (o famoso INPA) escreveu um artigo intitulado "Matemática contribui com 18% do PIB da França" [13]. E o artigo começa assim: " O imperador Napoleão Bonaparte escreveu que "o avanço e a perfeição da matemática estão intimamente ligados à prosperidade do Estado". Sua visão moldou a França moderna, e seus compatriotas colhem os benefícios até hoje". Marcelo discorre, então, sobre uma pesquisa recente realizada com o objetivo de aferir qual o impacto econômico da matemática na França, país que detém 22% dos ganhadores da Medalha Fields (o prêmio Nobel da matemática). O resultado é impactante: cerca de 281 bilhões de euros (os tais 18% do PIB) são gerados por empregos que demandam grande conhecimento de matemática.

Esses dados mostram a importância de levarmos a ciência até as populações mais carentes e do relevante trabalho necessário para atrair as crianças e os jovens para as áreas científicas. As mulheres, como mencionado acima, são mais de 50% da população brasileira. Portanto, para além da justiça social, diversidade, equidade e inclusão são responsáveis por aumentar a criatividade e contribuir para a competitividade positiva. Estudos mostram que diversidade de gênero torna a ciência mais diversa, mais eficiente, mas produtiva, melhor [14].

Como a diversidade é vista como um mecanismo eficiente de inovação e desenvolvimento, o progresso da ciência e da tecnologia pode se beneficiar da diversidade e, reciprocamente, pode exercer um *feedback* positivo ao mudar a imagem observada da desigualdade. No entanto, a ciência reproduz padrões sociais e o *Global Gender Gap Report* 2023, [15] produzido pelo *World economic Forum* , mostra que o Brasil ocupa a nada honrosa 57ª posição quando o assunto é equidade de gênero e melhorou substancialmente no último ano porque estávamos em 94º lugar em 2022. Os principais indicadores utilizados nessa classificação são oportunidade e participação econômica das mulheres (vamos mal), alcance educacional (estamos muito bem), saúde e sobrevivência (estamos bem) e empoderamento político feminino (simplesmente, não existe no Brasil). Os primeiros países são Islândia, Noruega, Finlândia, Nova Zelândia, mas o oitavo lugar é da Namíbia e o décimo segundo de Ruanda. O que têm Namíbia e Ruanda que o Brasil não tem? Uma explicação singela foi a existência de guerras, que empurraram as mulheres para posições de poder e de definição de políticas internas. A

Namíbia conquistou a independência em 1990 e Ruanda passou por uma guerra civil entre 1990 e 1994. O Brasil pode trilhar caminhos menos violentos para melhorar a equidade e, como já mencionei acima, a atração de mulheres para a ciência é um deles.

Para entendermos melhor os dados da ciência brasileira, em setembro de 2023, o CNPq lançou o Painel de Fomento em Ciência, Tecnologia e Inovação [16], concebido e implementado pela Diretoria de Análise de Resultados e Soluções Digitais (DASD), com o propósito de contribuir para as ações de monitoramento e avaliação da política científica e tecnológica realizada pelo CNPq, além de se caracterizar como importante ferramenta de transparência e prestação de contas à sociedade. Desde então, outros quatro painéis foram disponibilizados, sendo eles o Mapa de Fomento em Ciência, Tecnologia e Inovação – Bolsas e Projetos Vigentes [17], o Painel de Demanda e Atendimento de bolsas e auxílios [18], o Painel Lattes [19], que reúne dados extraídos dos currículos dos quase um milhão de mestres e doutores, cadastrados na Plataforma Lattes, e que atualizaram seus currículos nos últimos 5 anos e o Painel de Chamadas de Bolsas de Produtividade – PQ (de 2013 a 2023) [20]. Esses painéis, isoladamente ou em conjunto, permitem uma séria de análises com relação ao financiamento e a produção de mulheres cientistas. Parte dessas análises podem ser encontradas em [21], mas alguns pontos merecem ser também aqui destacados. Cabe salientar que o CNPq tem adotado várias medidas buscando garantir a maior participação das mulheres na ciência, como a possibilidade de prorrogação de várias modalidades de bolsas em caso de parto e adoção e a inserção no Currículo Lattes do campo Licença Maternidade e Adoção, além de recomendar aos comitês assessores do CNPq a admissão de medidas efetivas para a correção das possíveis lacunas de gênero e étnico-raciais existentes nas avaliações das chamadas. Por outro lado, a própria composição dos comitês assessores, responsáveis por assessorar o CNPq na avaliação de projetos e programas relativos à sua área de competência, vem paulatinamente alcançado equidade de gênero entre os seus participantes, uma iniciativa apoiada fortemente pelo Conselho Deliberativo do CNPq, como se pode acompanhar no gráfico 1 de [21] e no Painel onde constam os membros dos Comitês Assessores de 2023 e 2024 [22]. Deste último, observa-se que o percentual de pessoas do sexo feminino passou de 47,35% em 2023 a 51,93% em 2024. No entanto, alguns Comitês, geralmente ligados às ciências exatas e engenharias, ainda apresentam baixa representatividade, num reflexo do que ocorre na academia, tanto no Brasil, como no mundo.

Se na sociedade e na academia, a situação já é bastante complicada, quando, então, deveremos ter equidade de gênero na física, uma área tão masculinizada? Segundo um estudo recente, caso o passo atual não mude, apenas em 2158 a proporção de mulheres autoras de pesquisas na área se equipará a de homens [23]. E o que é mais chocante nesse quadro é o fato de uma grande parcela de homens físicos brancos não ter a menor noção de que a falta de diversidade e os fatores que afastam as mulheres da carreira devem ser encarados também por eles. Muitos nem sequer percebem que desempenham um papel importante no quadro de desigualdade [24].

## 3. Ações de combate à desigualdade

Dessa forma, grande parte das ações afirmativas na direção de uma maior diversidade, equidade e inclusão ficam por conta das próprias mulheres e de organismos ligados à ciência. A importância de prêmios que visem dar visibilidade ao trabalho feminino é inegável porque as coloca como modelos possíveis de serem alcançados, o que serve de estímulo às meninas e jovens mulheres. Nessa linha, há várias iniciativas que merecem ser citadas, mas vou me restringir a algumas:

- **Prêmio Mulheres e Ciência,** criado em 2024 pelo Conselho Nacional de Ciência e Tecnologia (**CNPq**) [25], com a intenção de promover a diversidade, a pluralidade e a participação de mulheres nas carreiras de ciência, tecnologia e inovação.

- **Prêmio Carolina Nemes**, outorgado pela Sociedade Brasileira de Física (**SBF**) [26] para mulheres físicas em início da carreira cujo trabalho de pesquisa tenha contribuído de forma significativa para o avanço da física ou do ensino de física no país. O prêmio visa reconhecer contribuições de mulheres para o desenvolvimento da física brasileira, bem como contribuir para diminuir a desigualdade de gênero na física.
- **Prêmio Carolina Bori Ciência e Mulher** oferecido pela Sociedade Brasileira para o Progresso da Ciência (**SBPC**) [27] é uma homenagem da SBPC às cientistas brasileiras destacadas e às futuras cientistas brasileiras de notório talento, que leva o nome de sua primeira presidente mulher, Carolina Martuscelli Bori.
- **Programa L'Oreal-UNESCO-ABC** para Mulheres na Ciência, oferecido pela Academia Brasileira de Ciências (**ABC**), em parceria com a **L'Oréal** e a **Unesco** [28], premia anualmente com uma Bolsa Auxílio (*Grant*) jovens doutoras brasileiras com projetos científicos de alto mérito a serem desenvolvidos durante 12 meses em instituições nacionais.
- **Prêmio Propesq-Mulheres na Ciência,** criado pela **UFSC** [29] com o propósito de estimular, valorizar e dar visibilidade às mulheres da UFSC que fazem pesquisas científicas, tecnológicas e inovadoras, divulgando-as amplamente.

Outra linha de ação visa estimular discussões que envolvem questões afetas às mulheres cientistas. Em 2015, a Assembleia Geral das Nações Unidas estabeleceu o dia 11 de fevereiro como o **Dia Internacional das Mulheres e Meninas na Ciência** [30], data que a UNESCO vem celebrando desde então para fomentar a igualdade de gênero e o papel de mulheres e meninas na ciência, dando apoio a meninas e jovens mulheres, sua formação e suas habilidades. O objetivo é a atração de jovens para as áreas ditas STEM, uma vez que apenas 35% dos estudantes dessas áreas são do gênero feminino.

No Brasil, um projeto que ganhou grande notoriedade nos últimos anos, por sua constante combatividade é o **Parent in Science** [31], que surgiu com o intuito de levantar a discussão sobre a parentalidade dentro do universo da academia e da ciência.

Apesar das distintas ações com focos ligeiramente coincidentes mencionados acima, quando se fala de dar visibilidade e espaço de fala às mulheres que atuam na ciência e, em particular, na física, elas ainda são poucas. Nesse contexto e num momento durante o qual tanto as mulheres quanto a ciência recebiam críticas contundentes com a finalidade de desacreditá-las, iniciei o projeto **Mulheres na Ciência** [32].

## 4. Projeto Mulheres na Ciência

No Brasil, a alfabetização científica obtida no ambiente escolar é incipiente. Em países mais ricos e que valorizam mais o conhecimento científico, há museus de ciência que ajudam a cobrir essa lacuna. Aqui, há pouquíssimos espaços com esse propósito. Esses fatores resultam numa sociedade despreparada para lidar com situações quotidianas que dependem de um conhecimento científico mínimo. O letramento científico é de vital importância para entendermos melhor a segurança e o risco envolvidos nos fenômenos naturais e nas novas tecnologias que estão ao nosso alcance.

Existem na nossa sociedade estruturas muito bem organizadas e financiadas com o objetivo de ajudar a disseminação das pseudociências e das *fake news.* Tais estruturas são, muitas vezes, geradas pela submissão sistemática de poderes públicos a interesses privados, bem como pelo desconhecimento generalizado de como funciona a ciência. Consequentemente, essas estruturas impõem aos cientistas um esforço extra, para além de fazer ciência: o de comunicar o que é e o que não é ciência, trabalho nada simples para o qual os cientistas não estão treinados. Contribui para o quadro atual, o fato de que a ponte entre a ciência de fronteira realizada e a percepção social do conhecimento por ela gerado é difícil de ser construída. Nesse contexto, o envolvimento de jovens cientistas no projeto também serve de treino para que se envolvam com divulgação científica e a valorizem.

O projeto teve início com um canal aberto no *Youtube* em setembro de 2019, cujo objetivo foi trazer a público diversos tópicos científicos de forma descomplicada e objetiva por meio de filmes curtos produzidos e protagonizados por cientistas e estudantes mulheres. A ideia inicial era usar o canal para contribuir, ao mesmo tempo, com o letramento científico dos brasileiros, a divulgação científica de assuntos de ponta e lutar contra o falso estereótipo de gênero que enxerga mulheres como menos competentes do que os homens. Os roteiros são pensados e escritos por mulheres e os vídeos são também protagonizados por mulheres. Os filmes discutem desde a segurança envolvida na utilização do forno de micro-ondas e nos celulares até assuntos de saúde pública, como a luta contra as superbactérias e a falácia por trás das terapias quânticas.

Em abril de 2021, por sugestão de uma aluna de primeira fase do curso de física, o primeiro vídeo curto foi veiculado na rede TikTok e daí para a frente, seguiram-se postagens no Instagram (*posts* e *reels*) e, a partir de agosto de 2021, também no s*horts* do *YouTube*. O YouTube conta com 7.000 inscritos e vídeos que chegaram a mais de 20.000 visualizações, enquanto o TikTok possui 21.000 seguidores e postagens com mais de 400.000 visualizações. O fato do projeto dar visibilidade e voz a várias cientistas, de diversas áreas do conhecimento, não fideliza o público a uma imagem única, como é recorrente em canais com muitos seguidores. Essa vulnerabilidade é bastante óbvia, mas a ideia não é dar notoriedade a uma ou duas pessoas, mas manter o objetivo de divulgar ciência de qualidade e treinar possíveis futuras divulgadoras de ciência. Uma *playlist* é dedicada a Mulheres Incríveis [33], que se projetaram a partir de descobertas ou comportamentos excepcionais.

**5. Conclusão**

A grande disseminação de desinformação e de discursos de ódio nas mídias sociais agravada pela falta de letramento científico da população brasileira contribuem para um quadro assaz preocupante no que tange a falta de equidade de gênero. O quadro é bastante sério na sociedade, como mostram os estudos publicados no *Gender Gap Report*, mas é ainda mais complicado nas áreas acadêmicas conhecidas por STEM, uma vez que em grupos muito masculinizados, problemas de falta de equidade tendem a ser minimizados ou mesmo ignorados.

No entanto, apesar do inegável quadro de desigualdade e injustiça social ao qual estão expostas as mulheres em geral e as cientistas em especial, várias ações nacionais e internacionais têm surgido para dar-lhes voz, apoio e visibilidade.

A falta de equidade de gênero deve ser assunto recorrente nos ambientes acadêmicos e o enfrentamento dos obstáculos por grupos sub-representados precisa ser lembrado e confrontado, se quisermos caminhar para uma ciência brasileira mais justa e diversa.